\documentclass[12pt]{article}
 \usepackage{graphicx}
\usepackage{epsfig}
 \usepackage{psfig}
\begin{document}
\begin{center}
{\bf{Bose-Einstein Condensation in a uniformly accelerated frame}}\\
Sanchita Das$^{a)}$ and Somenath Chakraborty$^{b)}$\\
Department of Physics, Visva-Bharati, Santiniketan, India 731235\\
Email:$^{a)}$sanchita.vbphys@gmail.com\\
Email:$^{b)}$somenath.chakrabarty@visva-bharati.ac.in
\end{center}

\begin{center}
Abstract
\end{center}
In this article we have investigated the possibility of Bose-Einstein Condensation (BEC) in a 
frame undergoing uniform acceleration or in other wards, in Rindler space associated with the
uniformly accelerated frame. We have followed a very simple conventional technique generally 
used in text book level studies. It has been observed that the critical temperature for BEC 
increases with the increase in magnitude of acceleration of the frame. Typically the critical 
temperature in an accelerated frame is of the order of the Unruh temperature. Hence we have 
concluded that the increase in the magnitude of acceleration of the frame facilitates the
formation of condensed phase.

Since bosons do not obey Pauli exclusion principle, the total wave function for a system
consisting of a number of bosons is symmetric in nature. Then at low temperature, less than
a typical value known as the critical temperature for BEC, a substantial number of bosons
will go to the ground state energy level or  zero energy level \cite{R1,R2}. In a certain
sense, the BEC is akin to the familiar process of liquid-vapor phase transition.
Conceptually, however, the two process are very different. Unlike the liquid-vapor
transition, which occured in configuration space, the BEC takes place in momentum
space-described as a condensation in momentum space \cite{R3}. However with the critical
examination of the equation of states of non-BEC and BEC phases shows the first order nature
of the phase transition \cite{R3}. The term momentum space condensation is the thermodynamic
manifestation that the transition to BEC state occurs only because of the symmetric nature
of the wave function and not because of any inter particle interaction \cite{R4}. This is
the very reason why it takes so many years to verify it experimentally. Although it has been 
predicted long ago \cite{R1,R2}, only in the year 1995 the phenomenon was experimentally 
verified \cite{R5,R6}. The advancement of technology, in particular,the progress in low
temperature physics, including the associated cryogenic technology helps a lot to conduct
this experiment. Two experimental condensed matter physicist, Eric Cornell and Carl Wieman 
were awarded Nobel prize in the year 2001 to verify the formation of BEC. They have used a few 
hundreds of $Rb^{87}$ atoms, which are bosons. These atomic bosons were trapped in a harmonic 
potential \cite{R3}. They have obtained BEC at around 170 Nano K.

Let us now introduce in brief the concept of Rindler space. It is associated with a frame 
undergoing uniform accelerated motion with respect to some inertial frame \cite{R7,R8,R9,R10}. 
The Rindler space can very easily be shown to be flat in nature (curvature tensors are zero in 
this case). Therefore one can say that the Rindler space is a kind of flat Minkowski space where 
the uniform velocity of the moving frame (inertial in nature) is replaced by uniform acceleration 
(non inertial in nature). Now from the principle of  equivalence a frame undergoing acceleration 
in absence of gravity is equivalent to a frame at rest in presence of a gravitational field
of which the strength of gravitational field 
is exactly equal to the magnitude of the acceleration \cite{R10}. Therefore in the present situation 
it is equivalent to say that we are going to study BEC in a rest frame but in presence of a strong 
uniform background gravitational field.

Now it can very easily be shown that in the Rindler space the Hamiltonian of a particle of rest 
mass $m_0$ is given by \cite{R11,R12,R13,R14,R15} (see also \cite{R16,R17,R18})\\
\begin{equation}
H=\left(1+\frac{\alpha x}{c^2}\right) \left(p^2 c^2 +m_0^2 c^4\right)^{1/2}
\end{equation} \\
where it is assumed that the frame is undergoing accelerated motion along $x$-axis, which is of 
course arbitrary in nature, indicates that the space spanned on $x-y$ plane is isotropic. The 
acceleration or the gravitational field $\alpha$ is constant within a length parameter $x_0$ 
(say). Then  in our formalism we replace $x$ in the above expression by $x_0$. For the sake of 
convenience one can use without the loss of generality the non-relativistic approximation of the 
above equation, given by
\begin{equation}
H=\left(1+\frac{\alpha x}{c^2}\right) \left(\frac{p^2}{2m_0} +m_0 c^2\right)
\end{equation} 
At this point we should mention that in quantum mechanical picture the above Hamiltonian $H$ 
(both in relativistic and non relativistic scenario) is non-Hermitian. But is $PT$ symmetric 
in nature, where $P$ is the parity operator and $T$ is the time reversal operator \cite{R19}. 
As a consequence the eigen spectrum of $H$ is real in nature. Further, the effect of background 
gravitational field on the energy eigen value has come through the Rindler Hamiltonian $H$ 
containing the term $\alpha$. We have noticed that even in quantum mechanical studies in Rindler 
space the strong gravitational field affects the binding of the particles \cite{R17}. Now below 
the critical temperature $T_c$ for the transition to BEC one can write
\begin{equation}
n=n_{p \neq0}+ n_{p=0}
\end{equation} 
where $n$ is the total number of bosons per unit volume and on the right hand side, the first 
term is the total number of normal bosons per unit volume and the second term is the corresponding 
quantity per unit volume in the BEC phase. In the usual case
\begin{equation}
n=\frac{1}{\lambda^3}g_{3/2} (z) +\frac{z}{1-z}\times\frac{1}{V}
\end{equation} 
where $\lambda={h}/{(2\pi m k T)^{1/2}}$, the thermal de-Broglie wavelength, $z$ is the fugacity,
$V=S_{yz}dx$ the volume of a small cylinder of cross sectional area $S_{xy}$ with the corresponding length element $dx$
and the function $g_{3/2} (z)$ is given by the infinite series
\begin{equation}
g_{3/2} (z)=\sum_{l=1}^{\infty} \frac{z^l}{l^{3/2}}
\end{equation}
Here the volume element $V$ is completely arbitrary in which the field $\alpha$ is constant, however, we are not
using this expression explicitly in our analysis. Therefore the arbitrariness is not going to affect our conclusion.
The function $g_{3/2} (z)$ is a monotonically increasing function of $z$ and is bounded within 
the domain $0\leq z \leq 1$. At $z=1$, $g_{3/2} (1)=2.612$, beyond which it diverges to $+\infty$.
Therefore at $T=T_c$, the critical condition \cite{R3}\\
\begin{equation}
n=\frac{1}{\lambda^3}g_{3/2} (1)
\end{equation} 
gives the critical temperature for BEC. Obviously at $T_c$, the chemical potential $\mu$ for the 
Bose particles is zero or in the relativistic scenario is $m_0 c^2$, the rest mass energy. 
Therefore it is also quite clear that in the Rindler space, the value of $\mu$ and $T_c$ will be 
completely different from that of the usual scenario. In Rindler space at $T=T_c^{(\alpha)}$ (say),
where  $T_c^{(\alpha)}$ is the modified critical temperature in the Rindler space and  
\begin{equation}
\mu= \left(1+\frac{\alpha x_0}{c^2}\right) m_0 c^2 =u(\alpha) m_0 c^2
\end{equation}
is the modified form of chemical potential when the system is under going an uniform acceleration 
$\alpha$. Obviously, the value of the chemical potential is large enough in the Rindler space if 
the magnitude  of the acceleration is also quite high \cite{R14}. Of course here also one has to 
consider 
 $z=1$ and then the total number of particles per unit volume is given by
 \begin{equation} 
 n=\frac{1}{u^{3/2}\lambda_{\alpha}^3}g_{3/2} (1)
 \end{equation}
where $u=\left(1+\frac{\alpha x_0}{c^2}\right)$ and $\lambda_\alpha$ is the thermal de
Broglie wavelength in the Rindler space. It is quite obvious that we get back the usual
solution if $\alpha=0$, i.e., in the inertial frame or $\alpha x_0/c^2 \ll1$.
Since $n$ is fixed (there is also no change in volume), we have from eqn.(8) (see eqn.(41) of 
\cite{R14})
\begin{equation}
T_c^{(\alpha)} =u(\alpha) T_c =\left(1+\frac{\alpha x_0}{c^2} \right) T_c
\end{equation}
This is the mathematical relation connecting the critical temperatures for BEC in the two 
different physical scenarios.Therefore in our simplified picture, the critical temperature for 
BEC transition will be quite high if the quantity ${\alpha x_0}/{c^2}\gg1$. Which may happen  
because of the  large enough  value of $\alpha$, the strength of back ground gravitational field.
Assuming a typical value $x_0=1 KM=10^5cm $ and $T_c=10^{-7}$K, We have
\begin{equation}
T_c^{(\alpha)}\approx \left(10^{-7} +10^{-23}\alpha\right)K
\end{equation}
where the uniform acceleration of the frame is expressed in the unit cm/s$^2$.
Therefore to affect $T_c$, the minimum value of uniform acceleration or the constant 
gravitational field must be $\approx 10^{16} cm/s^2$ and then only, the second part on the 
right hand side of eqn.(10) will be of the same order of magnitude as that of the first term. 
Let us now compare the second term with the well known Unruh temperature \cite{R20,R21}, given by 
\begin{equation}
T_U= \frac{\hbar \alpha}{2\pi c k_B}
\end{equation} 
The Unruh effect is associated with the  quantum mechanical interaction between an accelerated 
observer and the inertial vacuum. Since the accelerated observer carries so much extra energy 
that it can transfer some of it to the inertial vacuum at the time of interaction and as a result 
the 
inertial vacuum will no longer remain vacuum. The inertial vacuum will be excited and emits 
radiations and particles. One can say that the vacuum state will become warmer to the
accelerated observer. However, it will 
remain vacuum to an inertial observer at rest or in motion with uniform velocity. The expression 
for Unruh temperature is \cite{R20,R21}
\begin{equation}
T_U=\frac{\hbar \alpha}{2\pi c k_B}=4\times10^{-23} \alpha /[cm/s^2][K]
\end{equation}
Therefore the second term on the right hand side of eqn.(10) is $\approx T_U$. Hence we may write 
\begin{equation}
T_c^\alpha=T_c+ T_U
\end{equation}
If $T_U\gg T_c$, we can conclude that the critical temperature $T_c^{(\alpha)}$ for BEC
transition, in Rindler space is  $\approx T_U$ \cite{R22}. Therefore we may conclude that in
our simplified picture the uniform acceleration of the moving frame containing the Bose gas,
the critical temperature for BEC can be quite high when observed from an inertial frame. It
is shown in \cite{R22} that there exist a critical acceleration above which there is no BEC
whereas it begins to appear in the accelerating frame when the acceleration is gradually
decrease. We have noticed that in our formalism the chemical potential increases
monotonically with the increase in the uniform acceleration of the moving frame. Which is
consistent with the result obtained in \cite{R22}. In \cite{R23} the authors have obtained
negative thermal-like correction associated with acceleration of the frame. They have 
used the thermo-field dynamics and the effect of uniform acceleration of the
moving frame and gave an explanation for this negative sign and shown the increase in the
fractional abundance of condensed phase with larger acceleration. Indirectly speaking our 
formalism is 
also consistent with the work presented in \cite{R23}. In our model the critical temperature 
$T_c^{(\alpha)} \approx T_U$. Therefore if the temperature of the Bose system in the uniformly 
accelerated frame is decreased continuously i.e., $T$ becomes more and more less than the 
corresponding critical temperature, then it is quite possible that almost all the bosons 
will go to the condensed phase much before absolute zero temperature. To illustrate graphically, in fig.(1) we have
plotted 
\[
\frac{N_0}{N}=1-\frac{T^3}{{T_c^{(\alpha)}}^3}
\]
We have considered just for the sake of illustration $T_c^{(\alpha)}={10^7}^o$K, the core temperature of sun and
vary $T~~(\leq T_c^{(\alpha)})$ from ${10^{-7}}^o$K, the approximate value of the critical temperature for BEC in
the laboratory to $T_c^{(\alpha)}$, the critical temperature for BEC in the Rindler space with the value of
acceleration $\alpha \sim 10^{30}$cm/s$^2$, which is also the corresponding Unruh temperature. It is quite obvious
from the nature of the curve that all the bosons go to their ground state $u(\alpha)m_0$ much before $T=0$.
Therefore we may conclude that the increase in uniform acceleration of the frame or equivalently, the constant
gravitational field of the rest frame facilitates the BEC process.  

\begin{figure}[ht]
\psfig{figure=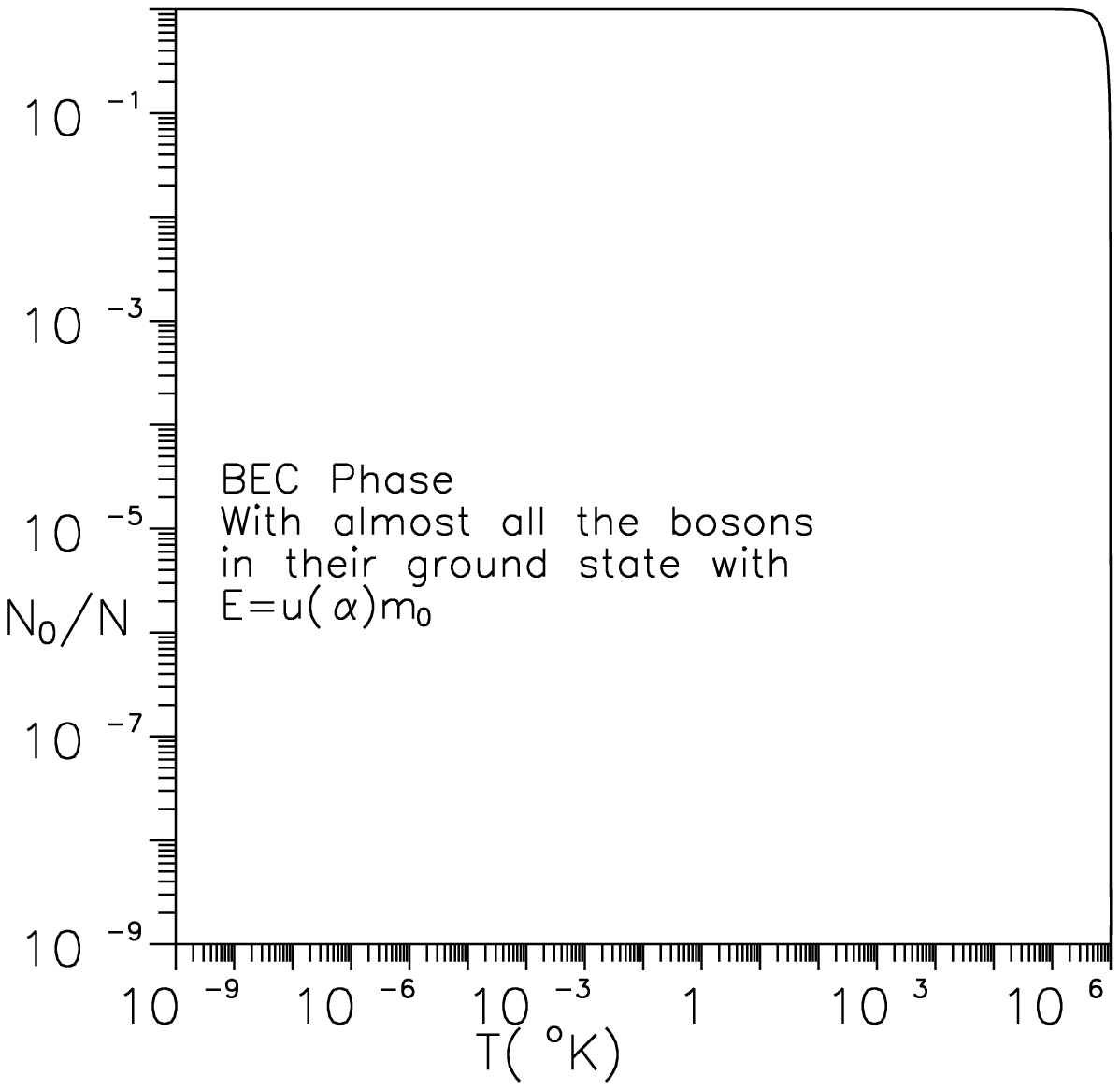,height=0.7\linewidth} 
\caption{Phase diagram for BEC in Rindler space (an illustration only)}
\end{figure}
\end{document}